\documentstyle[preprint,aps,epsf,amsmath,amssymb]{revtex}
\def\JHEP #1 #2 #3 {J. High Energy Phys. {\bf #1}, #2 (#3)}
\def\PR #1 #2 #3 {Phys.~Rev.~{\bf #1}, #2 (#3)}
\def\PRL #1 #2 #3 {Phys.~Rev.~Lett.~{\bf #1}, #2 (#3)}
\def\PRD #1 #2 #3 {Phys.~Rev.~D~{\bf #1}, #2 (#3)}
\def\PLB #1 #2 #3 {Phys.~Lett.~B~{\bf #1}, #2 (#3)}
\def\NPB #1 #2 #3 {Nucl.~Phys.~{\bf B#1}, #2 (#3)}
\def\RMP #1 #2 #3 {Rev.~Mod.~Phys.~{\bf #1}, #2 (#3)}
\def\ZPC #1 #2 #3 {Z.~Phys.~C~{\bf #1}, #2 (#3)}

\def\RPV {$R_p$-violating }
\def\RPC {$R_p$-conserving }
\def\lpp {\lambda^{\prime\prime} }

\begin{document}
\draft
\preprint{ANL-HEP-PR-00-062}

\title{Direct probes of $R$-parity-violating supersymmetric couplings \\
via single-top-squark production}
\author{Edmond L. Berger, B. W. Harris, and Z. Sullivan}
\address{High Energy Physics Division,
             Argonne National Laboratory,
             Argonne, Illinois 60439 }
\date{December 14, 2000}
\maketitle

\begin{abstract}
We study the $s$-channel production of a {\it single} top squark in hadron
collisions through an $R$-parity-violating mechanism, examining in detail
the case in which the squark decays through an $R$-parity-conserving
process into a bottom quark, a lepton, and missing energy.  We show that
the top squark can be discovered if its mass is less than 400 GeV, or that
the current bound on the size of the $R$-parity-violating couplings can be
reduced by up to one order of magnitude with existing data and by two
orders of magnitude at the forthcoming run~II of the Fermilab Tevatron.
\end{abstract}

\vspace*{5ex}
\section{Introduction}
\label{intro}

In supersymmetric extensions of the standard model, particles may be
assigned a new quantum number called $R$ parity ($R_p$)~\cite{farrar}.  The
particles of the standard model are $R_p$ even, and their corresponding
superpartners are $R_p$ odd.  If $R_p$ is conserved, as is often assumed,
superpartners must be produced in pairs, each of which decays to a final
state that includes at least one stable lightest supersymmetric particle
(LSP).  If $R_p$ is not conserved, then many of the experimental signatures
of supersymmetry that are usually studied may not be observable.  Hence, it
is important to determine whether $R_p$ violation occurs and, if so,
whether the effects are significant in collider experiments.  Current
bounds on possible $R_p$-violating couplings, obtained principally from
quantum corrections to particle decays and neutral meson mixing, are
relatively restrictive for the first generation of quarks and leptons but
are much less so for states of the second and third
generations~\cite{dreiner,llt}.

Searches for $R_p$ violation often focus on the decays of superpartners
under the assumption that they are produced in pairs via an \RPC
interaction~\cite{dreiner}.  However, squarks and gluinos are generally
predicted to be relatively heavy, and therefore their pair-production cross
sections incur large phase space suppressions at Fermilab Tevatron
energies.

In this paper, we expand and improve upon our previous
proposal~\cite{bhsletter} to probe \RPV couplings by searching for the
$s$-channel production of a {\it single} squark through an $R_p$-violating
mechanism~\cite{savas}.  The greater phase space afforded by the production
of one heavy particle, instead of two, will allow searches to reach much
higher masses than those based on traditional squark pair-production.
Further, since the $R$-parity violation occurs in production, the cross
section will provide a direct measure of the individual \RPV couplings,
rather than of products of different \RPV couplings.

In Sec.~\ref{superpot}, we describe the baryon-number-violating portion of
the $R_p$-violating Lagrangian and establish our notation and conventions.
The cross section for pair-production of top squarks is small compared to
that for the process $q q^{\prime}\to \widetilde{t}_1$ that can proceed
through \RPV couplings.  In order to capitalize on the enhanced rate for
single production it is necessary to examine the decay products of the top
squarks.  We address in Sec.~\ref{rpvdecay} the possibility that decay will
occur through an $R_p$-violating process into two hadronic jets and show
that this rate is far below that for standard quantum chromodynamics (QCD)
production of jet pairs.

In Sec.~\ref{rpcdecay}, we study observability of single top squarks by
focusing on the clean $R_p$-conserving decay, $\widetilde{t}_1 \rightarrow
b \widetilde{\chi}^+_1$, with $\widetilde{\chi}^+_1 \rightarrow l^+ + \nu_l
+\widetilde{\chi}^0_1$.  Here, $l$ is an electron or muon, and the
$\widetilde{\chi}^+_1$ and $\widetilde{\chi}^0_1$ are the lowest-mass
chargino and neutralino states of the minimal supersymmetric standard model
(MSSM)~\cite{mssm}.  We introduce a method to extract a top-squark mass
peak from the standard model background, principally $W +$~hadronic jets.
To study the effects of parton showering, hadronization, and
color-recombination, we simulate the signal with the {\sc
{\sc HERWIG}}~\cite{HERWIG} Monte Carlo event generation program.  The response
of a typical experimental detector is modeled using the full {\sc SHW} detector
simulation package~\cite{SHW}.  We explore a wide range of the parameter
space of the minimal supergravity (MSUGRA) model of supersymmetry
breaking~\cite{sugra}.  Each of these improvements renders the analysis
more realistic and strengthens the conclusions reached in our earlier work
\cite{bhsletter}.

Our conclusions are stated in Sec.~\ref{summary}.  For top squarks with
mass $m_{\widetilde{t}_1} < 400$ GeV, we show that discovery at the level
of five standard deviations ($5\sigma$) is possible at $\sqrt{S} = 2$ TeV
with an integrated luminosity of 2~fb$^{-1}$, provided that the
$R_p$-violating couplings $\lambda^{\prime\prime} >$ 0.02 --
0.1. Otherwise, a $95\%$ confidence-level exclusion limit can be set for
$\lambda^{\prime\prime}$ as a function of the mass up to about 500 GeV.
For the lower integrated luminosity and energy of the existing run~I data,
values of $\lambda^{\prime\prime} >$ 0.05--0.2 can be excluded at the
$95\%$ confidence level if $m_{\widetilde{t}_1}=$ 165--350 GeV.  These
limits would constitute a significant improvement over the current $95\%$
confidence-level upper bounds of $\lambda^{\prime\prime} \lesssim 1$ for
squarks of the third generation~\cite{dreiner,llt}. With the large increase
in cross section at the CERN Large Hadron Collider (LHC), the discovery
region and limits may be improved by another order of magnitude in $\lpp$
and to top-squark masses approaching 1 TeV.  The discovery reaches, or
exclusion limits, at the Tevatron and LHC depend on the value of the
top-squark mass, largely independent of other supersymmetric parameters.

A lower limit of about 165 GeV on the accessible value of the top-squark
mass via our method is determined by the selections made on the $b$ jet and
lepton momenta in order to achieve a satisfactory signal to background
ratio.  An upper bound of about 550 GeV is set by the event rate.  To
explore values of the top-squark mass well below that of the top quark, it
appears necessary to combine results of our analysis with those based on
pair production of top squarks.

For the values of the $R$-parity-violating couplings of interest to us, the
$R$-parity-violating width of the $\widetilde{\chi}^0_1$ remains negligible
and for all practical purposes the $\widetilde{\chi}^0_1$ LSP is stable.
We leave to future work the exploration of larger values of
$R$-parity-violating couplings and consequent decays of the
$\widetilde{\chi}^0_1$ within the fiducial region of detectors.


\section{MSSM Superpotential and Notation}
\label{superpot}

In general it is possible to have $R_p$-violating contributions to the
MSSM superpotential that violate baryon- or lepton-number.
However, limits on the proton decay rate severely restrict their
simultaneous presence.  In this paper, we assume the existence of only a
baryon-number-violating coupling of the
form~\cite{weinberg}
\begin{equation}
{\cal W}_{\not \! R_p} = \lambda_{ijk}^{\prime\prime}U_i^cD_j^cD_k^c \; .
\end{equation}
Here, $U^c_i$ and $D^c_i$ are right-handed-quark singlet chiral
superfields, $i,j,k$ are generation indices, and $c$ denotes charge
conjugation.  This form of the superpotential fixes our choice of
normalization of $\lambda_{ijk}^{\prime\prime}$.  It is equal to
$\lambda_H^{\prime\prime}/2$, where $\lambda_H^{\prime\prime}$ is the
coupling used in {\sc HERWIG}~\cite{HERWIG}.

In four-component Dirac notation, the Lagrangian that follows from this
superpotential term is
\begin{eqnarray}
{\cal L}_{\lambda^{\prime\prime}} & = & -2 \epsilon^{\alpha \beta \gamma}
\lambda^{\prime\prime}_{ijk}
\left[ \widetilde u_{R i \alpha} \overline{d^c_{j \beta}} P_R d_{k \gamma}
  +    \widetilde d_{R j \beta} \overline{u^c_{i \alpha}} P_R d_{k \gamma}
\right. \nonumber \\
     && \hspace*{6em} + \left.
     \widetilde d_{R k \gamma} \overline{u^c_{i \alpha}} P_R d_{j \beta}
\right] + h.c. \; ,
\end{eqnarray}
where $j < k$.

For production of a right-handed top squark via an $s$-channel subprocess
$\overline{d}_j\overline{d}_k \rightarrow \widetilde{t}_{R}$ the relevant
couplings are $\lambda^{\prime\prime}_{312}$,
$\lambda^{\prime\prime}_{313}$, and $\lambda^{\prime\prime}_{323}$.  The
most direct limits on these couplings come from the measurement of $R_l$,
the partial decay width of the $Z$ boson to hadrons over its partial decay
width to leptons.  The analysis of $R_l$ provides $95\%$ confidence-level
upper bounds of $\lambda^{\prime\prime}_{3jk} \lesssim
1$~\cite{dreiner,llt}.  The Bayesian limits,\footnote{In Ref.~\cite{llt},
it is stated that a classical statistical analysis rules out the
$\lpp_{3j3}$ couplings at the $2\sigma$ level, and places a bound of
$\lpp_{312} < 2.7$.  However, we do not agree that the analysis of
Ref.~\cite{llt} places a significant bound on $\lpp_{3j3}$.} given as a
function of mass in Ref.~\cite{llt}, are approximately
$\lambda^{\prime\prime}_{312} < 1.52 + 0.425\times(m_{\widetilde{t}_1}/100
{\rm \; GeV})$, and $\lambda^{\prime\prime}_{313,323} < 0.585 + 0.162\times
(m_{\widetilde{t}_1}/100 {\rm \; GeV})$.

Subsequent to our Letter~\cite{bhsletter}, it was argued that rare $B^+$
decays and $K^0$--$\overline{K}^0$ mixing constrain
$\lambda^{\prime\prime}_{3jk}$ strongly~\cite{slavich}.  The analysis of
these decays {\it indirectly} constrains {\it products} of different
$\lambda^{\prime\prime}_{3jk}$, whereas the method we propose may be used
to {\it directly probe each coupling independently}.

\begin{figure}[tb]
\begin{center}
\epsfxsize= 1.75in 
\leavevmode
\epsfbox{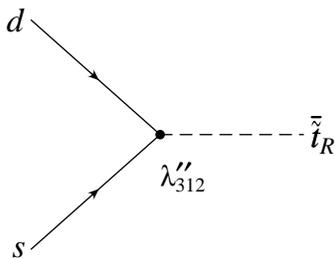}
\end{center}
\caption{Feynman diagram for the $s$-channel production of a single
top squark via $\lambda^{\prime\prime}_{3jk}$.}
\label{fig_feynprod}
\end{figure}

Shown in Fig.~\ref{fig_feynprod} is the Feynman diagram for $s$-channel
production of a single top squark via the partonic subprocess $d + s
\rightarrow \overline{\widetilde{t}}_R$.  The right-handed top-squark
interaction state is related to the mass eigenstates by $\widetilde{t}_R =
-\widetilde{t}_1 \sin\theta_{\widetilde{t}} + \widetilde{t}_2
\cos\theta_{\widetilde{t}}$.  For simplicity of notation, and motivated by
models where $\sin\theta_{\widetilde{t}}\approx 1$, we present results in
terms of a measurement of $\widetilde{t}_1$.  In general, the results are
valid for whichever mass eigenstates contain some amount of the
right-handed top squark.

The color- and spin-averaged cross section for inclusive $\widetilde{t}_1$
production is
\begin{equation}
\sigma =
\frac{2\pi}{3S} \sin^2\!\theta_{\widetilde{t}}\sum_{j<k}
|\lambda^{\prime\prime}_{3jk}|^2 \Phi_{jk}(\tau) \; ,
\label{sigma}
\end{equation}
where $S$ is the square of the hadronic center of mass energy, and $\tau=
m^2_{\widetilde{t}_1}/S$.  The integrated parton luminosity $\Phi_{jk}$
contains a convolution of the parton distribution functions (PDF's): $d
\otimes s$, $d \otimes b$, and $s \otimes b$, where $d$, $s$, and $b$ label
the PDF's of the down, strange, and bottom quarks, respectively.  The cross
section for $\widetilde{t}_2$ production is the same as above but with
$\sin^2\!\theta_{\widetilde{t}}$ replaced by
$\cos^2\!\theta_{\widetilde{t}}$.

\begin{figure}[tb]
\begin{center}
\epsfxsize= 3in 
\leavevmode
\epsfbox{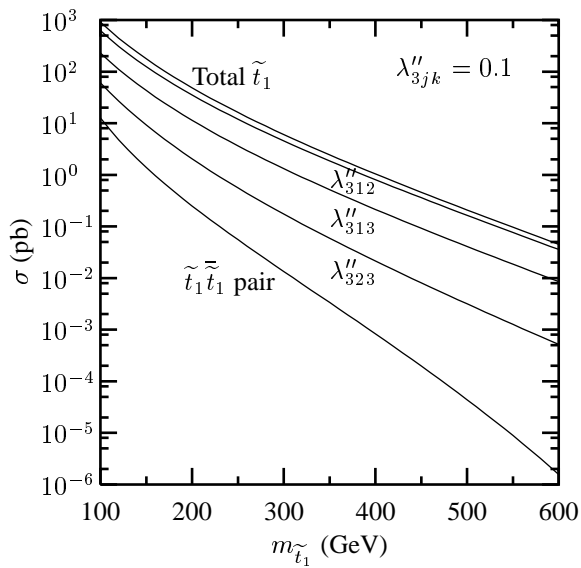}
\end{center}
\caption{Cross section versus top-squark mass $m_{\widetilde{t}_1}$
for $R$-parity-violating production of a single
top squark at run II of the Fermilab Tevatron ($\sqrt{S}=2$ TeV) with
$\lambda^{\prime\prime}_{3jk}=0.1$ compared with the
$R$-parity-conserving production cross section for top-squark pairs.}
\label{fig_compprod}
\end{figure}

The mass dependences of the next-to-leading order (NLO) cross sections for
$R_p$-violating single production~\cite{TPlehn} and $R_p$-conserving pair
production~\cite{zerwas} of top squarks differ significantly, as shown in
Fig.~\ref{fig_compprod}.  Results are presented for the sum of
$\widetilde{t}_1$ and $\overline{\widetilde{t}}_1$ production and couplings
$\lambda^{\prime\prime}_{3jk} = 0.1$.  Even if
$\lambda^{\prime\prime}_{3jk}$ is reduced to $0.01$, two orders of
magnitude below the current bound, the single-top-squark rate exceeds the
pair rate for $m_{\widetilde{t}_1}>100$ GeV.  The parton luminosities
determine that the contribution to the total cross section of the terms
proportional to $\lambda^{\prime\prime}_{312}:
\lambda^{\prime\prime}_{313}: \lambda^{\prime\prime}_{323}$ is about
$0.72:0.24:0.04$ at $m_{\widetilde{t}_1}=200$ GeV.

In order to simplify the explanation of some of the general features of
this study, we use the notation $\lambda^{\prime\prime} \equiv
\lambda^{\prime\prime}_{312} = \lambda^{\prime\prime}_{313} =
\lambda^{\prime\prime}_{323}$.  However, our method is identical even if
only one of the three couplings is non-zero.  The results for independent
couplings are related by a simple functional form and are presented below.
Since the cross section in Eq.~(\ref{sigma}) is proportional to the square
of the product $\lambda^{\prime\prime}_{3jk}
\sin\theta_{\widetilde{t}}$, a model-independent analysis of the data
limits the product.  Correspondingly, limits for $\widetilde{t}_2$ are
really on $\lambda^{\prime\prime}_{3jk}
\cos\theta_{\widetilde{t}}$.  In Sec.~\ref{rpcdecay}, we present
minimal supergravity models of supersymmetry breaking in which the mixing
angle $\theta_{\widetilde{t}}$ is uniquely determined by the model
parameters.  Within the context of these models, limits are placed on the
actual $\lpp_{3jk}$ couplings.


\section{$R$-Parity-Violating Decays}
\label{rpvdecay}

In the $R_p$-violating MSSM, the right-handed up-type squark
$\widetilde{u}_R$ can decay into quark pairs $\widetilde{u}_R
\rightarrow \overline{d}_j +\overline{d}_k$ via the $\lpp$ couplings.
The decay width of the top-squark mass eigenstate into two jets is
\begin{equation}
\Gamma(\widetilde t_1 \rightarrow j j) =
\frac{m_{\widetilde t}}{2\pi} \sin^2\!\theta_{\widetilde{t}}
\sum_{j<k} |\lambda^{\prime\prime}_{3jk}|^2 \; .
\end{equation}
If $\lpp \gtrsim 2$, the width of the top squark is greater that its mass.
This is true generally for {\it any} squark and {\it any}
baryon-number-violating coupling.  Hence, there is a practical upper limit
on $\lpp$ for the model to contain narrow resonances.

\begin{figure}[tb]
\begin{center}
\epsfxsize= 3.0in 
\leavevmode
\epsfbox{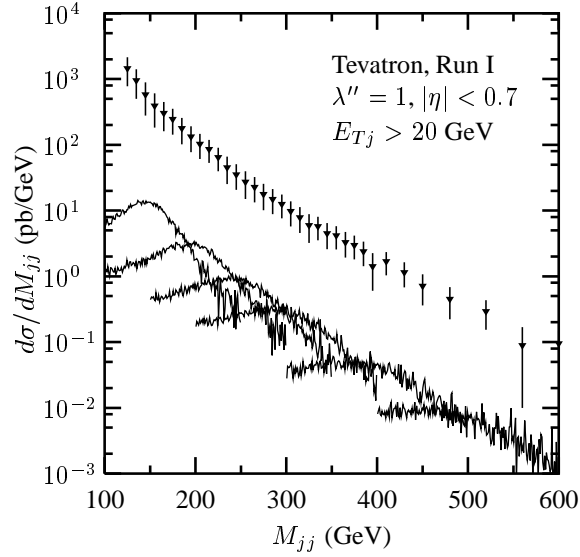}
\end{center}
\caption{Dijet mass distribution for the $R$-parity-violating decay
of the top squark at several masses (150, 200, 250, 300, 400, 500
GeV), compared with CDF dijet data from run I of the
Tevatron~\protect\cite{CDFdijet}.}
\label{fig_rpvdata}
\end{figure}

In Fig.~\ref{fig_rpvdata}, we present the dijet mass distribution that
results from the \RPV decays of singly-produced top squarks with $\lpp =
1$.  We simulate the signal for various masses with a matrix-element
calculation that uses {\sc HELAS}~\cite{HELAS} subroutines.  For
comparison, we also reproduce data published by the Collider Detector at
Fermilab (CDF) Collaboration~\cite{CDFdijet}.  We apply the same cuts in
our simulation as were used on the data.  Figure~\ref{fig_rpvdata} shows
that the signal to background ratio ($S/B$) in the dijet channel is less
than $1/100$ for any mass.  For $\lpp$ different from 1, this ratio is
reduced further to $S/B \lesssim 0.01\times(\lpp)^4/[(\lpp)^2+c]$, where
$c$ is a constant proportional to the \RPC width.  Because of the
overwhelming jet rate from standard strong interaction processes, we
confirm the expectation \cite{savas} that the dijet data can neither
exclude $R$-parity-violating production and decay of top squarks nor
provide useful bounds on $\lambda^{\prime\prime}$.


\section{$R$-Parity-Conserving Decays}
\label{rpcdecay}

In this section we focus on probes of $\lpp$ couplings through \RPC
decay modes of the top squark.  If kinematically allowed, a heavy top
squark decays into a bottom quark ($b$) and the lightest chargino
$\widetilde{\chi}_1^+$.  If $m_{\widetilde{t}_1} <
m_{\widetilde{\chi}_1^+} + m_b$, the decay would occur into a charm
quark and a neutralino via flavor changing loop diagrams~\cite{hikasa}.
For the top-squark masses of interest to us, and in the context of the
MSUGRA model in which we work, the decay $\widetilde{t}_1 \rightarrow b +
\widetilde{\chi}_1^+$ dominates.  Other channels open at large
$\widetilde{t}_1$ masses (e.g. decay into a top quark and the lightest
neutralino, $t+\widetilde{\chi}_1^0$), but their decay widths are less than
half of the \RPC decay width for the masses we consider.  The chargino
typically undergoes a three-body decay into a neutralino and the decay
products of a $W$ boson.  Since it is challenging to observe an all-jets
plus missing energy mode, we concentrate on the case where the final state
is a bottom quark plus a neutralino and an electron or muon from the
chargino decay.

\begin{figure}[tb]
\begin{center}
\epsfxsize= 3.0in 
\leavevmode
\epsfbox{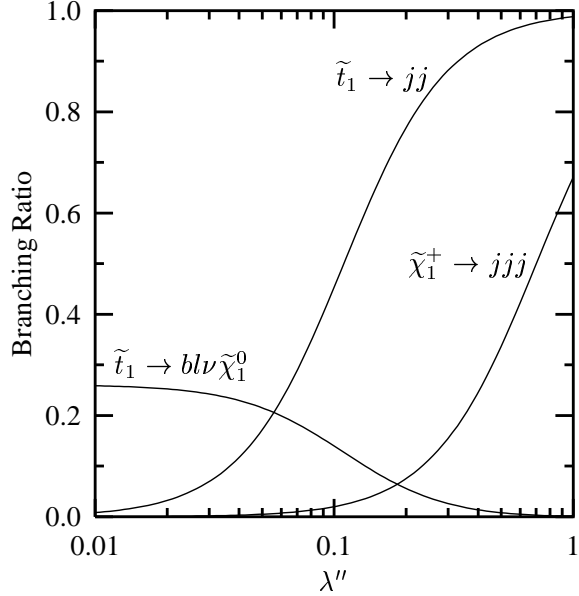}
\end{center}
\caption{Branching ratios for the top squark to decay into two jets and for
the chargino $\widetilde{\chi}_1^+$ to decay into three jets, both as a
function of the $R$-parity-violating coupling $\lambda^{\prime\prime}$.
Also shown is the branching ratio for the top squark to decay into
$b\, l \nu \widetilde{\chi}_1^0$.}
\label{fig_br}
\end{figure}

We show the branching fractions of the top squark into two jets and of the
cascade ($\widetilde{t}_1 \to b \widetilde{\chi}_1^+ \to b l^+ \nu_l
\widetilde{\chi}_1^0$) as a function of $\lpp$ in Fig.~\ref{fig_br}.  The
\RPV branching fraction of the chargino is also shown.  The \RPC decay
dominates for small $\lambda^{\prime\prime}$.  For both the top squark and
the chargino, the width into jets increases when $\lpp$ is large.  The
branching fraction into the signal mode is inversely proportional to
$(\lpp)^{2}$.  However, as seen in Eq.~(\ref{sigma}), the production cross
section is proportional to $(\lpp)^2$, and thus the entire cross section
behaves as

\begin{eqnarray}
\sigma(p\bar{p} \rightarrow b l^+ \nu_l \widetilde{\chi}_1^0) &=&
\sigma(p\bar{p} \rightarrow \widetilde{t}_1) \, \times \, {\rm BR}(
\widetilde{t}_1 \rightarrow b \widetilde{\chi}_1^+ ) \nonumber \\
&& \times {\rm BR}( \widetilde{\chi}_1^+ \rightarrow l^+ \nu_l
\widetilde{\chi}_1^0 ) \nonumber
\\ &\sim& {\lambda^{\prime\prime}}^2 \frac{1}{ {\lambda^{\prime\prime}}^2 +
a\Gamma_{\widetilde{t}_1}^{R_p}} \frac{1}{ {\lambda^{\prime\prime}}^2 +
b\Gamma_{\widetilde{\chi}_1^+}^{R_p}} \; .
\label{signalsigma}
\end{eqnarray}
In Eq.~(\ref{signalsigma}), $\Gamma_{\widetilde{t}_1}^{R_p}$ and
$\Gamma_{\widetilde{\chi}_1^+}^{R_p}$ are the \RPC partial decay
widths of the lightest top squark and chargino, respectively, and
$a$ and $b$ are appropriate proportionality factors.

In the presence of \RPV couplings, the neutralino is not stable.  There are
three decay scenarios that potentially give rise to distinct signatures in
the detector.  In the first scenario, only the $\lpp_{3jk}$ couplings are
larger than $10^{-4}$--$10^{-5}$.  In this case, the neutralino must decay
through an off-shell top quark and off-shell top squark into a five-body
final state.  A neutralino whose mass is $100$~GeV or less has decay
lifetime $c\tau > 100/(\lpp)^2$~m (see Fig.~26 of the first paper of
Ref.~\cite{dreiner}) and decays far outside the detector.  Hence, the
signature of the neutralino is simply missing energy in the detector.  In
this paper we address the case where the neutralino either decays outside
of the detector volume or leaves remnants that are too soft to identify.
The signal is a tagged $b$-jet, an electron or muon, and large missing
transverse energy.  The main backgrounds are single-top-{\it quark}
production, and $W+$jets processes that either contain a bottom quark or
have a jet that is mis-tagged as a $b$-jet.

If other baryon-number-violating couplings are large, then the neutralino
can decay inside the detector.  The signature of this second scenario
includes two or three extra jets that appear to come from the primary
vertex.  The modeling of the energy distributions of these jets depends on
explicit choices of some parameters.  In an interesting third possibility,
at least one of the first- or second-generation couplings,
$\lpp_{123},\lpp_{2jk}$, is greater than $10^{-3}$.  The decay may either
have a displaced vertex or occur in a calorimeter.  These possibilities
warrant further study.

\subsection{Simulation}
\label{simulation}

In order to include the effects of parton showering and hadronization, we
simulate the signal with the Monte Carlo program {\sc HERWIG}
$6.1$~\cite{HERWIG}, a version that includes \RPV interactions, and we use
the {\sc SHW} $2.3$~\cite{SHW} detector simulation package.  The {\sc SHW}
package provides a reasonably close approximation of the expected
acceptance of the upgraded CDF and D0 detectors to signal and background
processes. It determines what charged tracks and calorimeter energies the
detector would record and supplies information about the trigger and
reconstructed states such as electrons, muons, and hadronic jets, including
$b$- and $c$-jet tagging.

In addition to interfacing {\sc SHW} to {\sc HERWIG}, we make a few
modifications for this study.  We include the NLO $K$-factors calculated
in Ref.~\cite{TPlehn}.  A $k_T$ cluster algorithm~\cite{cluster} for
hadronic jets is added to provide an infrared-safe way of combining
calorimeter towers.  We use a $k_T$ cone size of 1, similar to a fixed cone
size of $\Delta R = 0.7$.  We also correct for the typical out-of-cone and
threshold energy losses in the jet reconstruction of about 4~GeV per jet.
After this correction, we find that the radiation, showering, and
color-reconnection modeled in {\sc HERWIG} 6.1 have little effect on the
distributions of the reconstructed objects with respect to the parton level
results presented in Ref.~\cite{bhsletter}.

We use impact-parameter tagging as defined in {\sc SHW} to tag $b$-jets and
to acquire the energy-dependent rate for charm quarks to fake bottom
quarks.  For a conservative estimate of the background, we choose the
mis-tag rate for light-quark jets to be the greater of $0.5\%$ or the
output of {\sc SHW}.  The event trigger is either an electron or muon that
passes the {\sc SHW} triggering condition listed in Table \ref{tab_acc},
typically a transverse energy $E_{T l} > 15$ GeV.  The missing transverse
energy ${\not \!E}_T$ is calculated for the entire detector (calorimeters
plus muon chambers) after energy corrections.  In order to stay above the
level of detector fluctuations we choose ${\not \!E}_T > 20$ GeV.  The
final significance is insensitive to the ${\not \!E}_T$ cut as long as it
is not raised above about $0.4\times m_{\widetilde{t}_1}$.

In Table~\ref{tab_acc}, we provide the geometric acceptance of a detector
as defined by {\sc SHW}.  We also list the minimum transverse energy used
to define identifiable objects ($b$-jets, leptons, or ``hard'' jets).  The
lepton-finding algorithm in {\sc SHW} is based on the CDF run I geometry
and thus covers a slightly smaller pseudorapidity region than used in our
original Letter (see Table I of Ref.~\cite{bhsletter}).  There is an
additional loss of $20\%$ of the leptons compared with our initial
assumptions due to modeling of detector efficiencies.  After these detector
effects and the slightly different choice of cuts are accounted for, the
full simulation is in complete agreement with the exact leading-order
matrix element calculation used in our Letter~\cite{bhsletter}.  We also
confirm that adding Eq.~(\ref{sigma}) to {\sc PYTHIA} 6.131 \cite{PYTHIA}
produces the same results, an independent confirmation that
color-recombination ambiguities, present for finite baryon-number-violating
couplings~\cite{dreiner2}, do not impact the results.

The backgrounds are modeled with tree-level matrix elements obtained from
MADGRAPH~\cite{MADGRAPH} convolved with leading-order CTEQ5L~\cite{CTEQ5}
parton distribution functions, at a hard scattering scale $\mu^2 =
\hat{s}$.  In order of importance, these backgrounds arise from production
and decay of the standard model processes $Wc$, with a charm quark $c$ that
is mistaken for a $b$; $Wj$, with a hadronic jet that mimics a $b$; $Wb\bar
b$; $Wc\bar c$; and single-top-{\it quark} production via $Wg$ fusion.  The
tagging efficiencies and geometry are the same as used in {\sc SHW}.  In an
experimental analysis, the $Wj$ background would be normalized by the data.
To simulate the resolution of the hadron calorimeter for background events,
we smear the jet energies with a Gaussian whose width is $\Delta E_j/E_j =
0.80/\sqrt{E_j}\oplus 0.05$ (added in quadrature).  To verify the accuracy
of our background estimation, we perform a full {\sc HERWIG} and {\sc SHW}
simulation for the $Wj$ background.  The relevant distributions match those
of the matrix element calculation to better than $10\%$.  As explained
below, numerical limits on the couplings are insensitive to small
uncertainties in the background.

The $b$ quark recoils against the chargino in the primary decay of the
top squark.  Thus the measured $E_T$ spectrum for the $b$-jet is peaked
near the kinematic limit
\begin{equation}
E_{T b}^{\rm max} = \frac{m^2_{\widetilde{t}_1} -
m^2_{\widetilde{\chi}_1^{+}} + m^2_b}{2 m_{\widetilde{t}_1}} \; .
\end{equation}
An estimate of the mass difference between the top squark and chargino may
be obtained if a prominent peak is found in the $E_{T b}$ spectrum.  If we
invert this equation, a kinematic constraint appears on the lowest
top-squark mass that may be probed with this method, $m_{\widetilde{t}_1}
> m_{\widetilde{\chi}_1^+} + E_{T b}^{\rm cut}$.

\begin{figure}[tb]
\begin{center}
\epsfxsize= 3.0in 
\leavevmode
\epsfbox{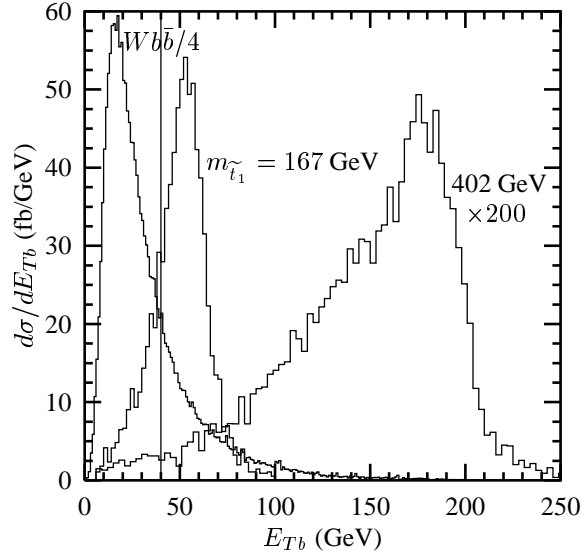}
\end{center}
\caption{The transverse energy $E_T$ spectrum (fb/GeV) of the tagged
$b$-jet for top squarks of mass 167 GeV and 402 GeV, and the $W b \bar b$
background at the Tevatron $\sqrt{S} = 2$ TeV.  We require
$E_{Tb} > 40$ GeV, marked by the vertical line.}
\label{fig_etb}
\end{figure}

In Fig.~\ref{fig_etb}, we show the $E_T$ spectrum for the $b$-jet, tagged
by the {\sc SHW} impact-parameter method, for two different top squark
masses, as well as the $W b \bar b$ background.  The background falls
rapidly as a function of $E_{Tb}$.  The $b$-jet from the single-top-quark
background peaks near 60 GeV, but it makes a small contribution after cuts.
We choose a cut of 40~GeV on the $E_T$ of the $b$-jet to ensure a
reasonable tagging efficiency (greater than $50\%$).  Variation of this
number between 30 GeV and 80 GeV has no effect on the significance of the
signal as long as the signal remains kinematically allowed.

\begin{figure}[tb]
\begin{center}
\epsfxsize= 3.0in 
\leavevmode
\epsfbox{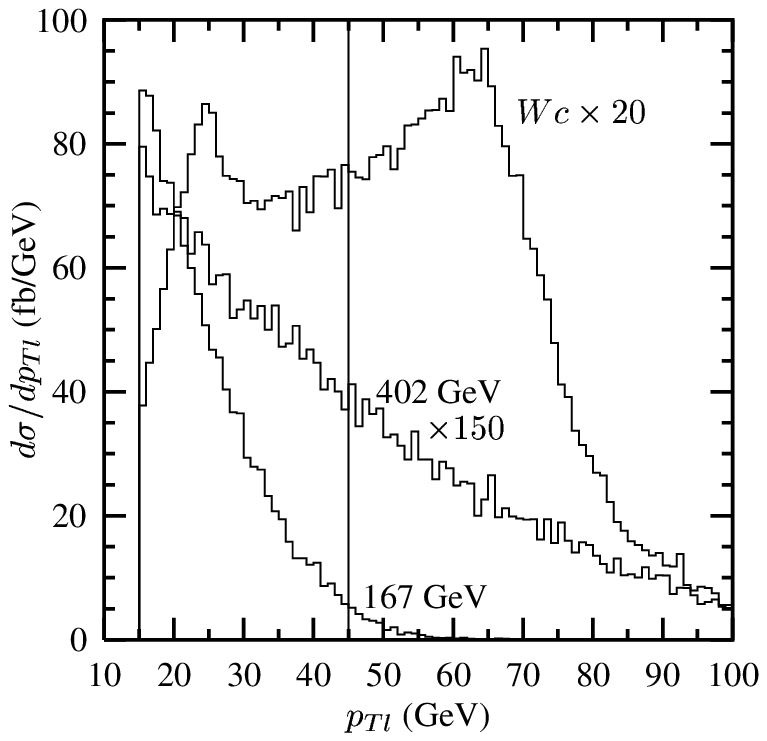}
\end{center}
\caption{The transverse momentum
$p_T$ spectrum (fb/GeV) of the tagged lepton for
top squarks of 167 GeV and 402 GeV, and the $W c$ background at the
Tevatron $\sqrt{S} = 2$ TeV. We require $p_{T l} < 45$ GeV
marked by the vertical line.}
\label{fig_ete}
\end{figure}

The lepton from chargino decay ($\widetilde{\chi}_1^+ \to l^+ \nu_l
\widetilde{\chi}_1^0$) tends to be soft since much of the energy goes into
the mass of the neutralino, and the rest is split by the three-body decay.
In contrast, the lepton in the background comes from decay of a real $W$.
It clusters around 40 GeV ($1/2$ of the $W$ mass) and has a broad rapidity
distribution.  After reconstruction, the remaining events are distributed
from 20 GeV to 70 GeV (see Fig.~\ref{fig_ete}).  Given the different shapes
of the $p_T$ spectra, we impose a hard ``lepton veto'' of 45 GeV to reduce
the background with very little effect on the signal.  This cut slightly
improves the reach of the analysis, but is most useful for increasing the
purity ($S/B$) of the Tevatron run I sample.

The extra hard jets in the $W b \bar b$, $W c \bar c$, and single-top-quark
backgrounds distinguish these processes from the signal.  Since extra jets
in the signal arise only in radiation from the hard process in 10--$20\%$
of the events, we consider a ``jet veto'' that removes any event containing
a second jet with $E_{T j} > 30$ GeV.  The main effect of this cut is to
improve the purity of the sample ($S/B$). The ``jet veto'' improves the
significance ($S/\sqrt{B}$) by a few percent at the Tevatron for high
$m_{\widetilde{t}_1}$, but decreases it by a few percent at the LHC for all
masses.  This cut is most useful if evidence of the signal is found and a
higher purity confirmation is desired.

\subsection{Results}
\label{results}

To obtain numerical results, we adopt a minimal supergravity
framework~\cite{sugra} to compute the masses of the top squark and its
decay products, the mixing angle $\theta_{\widetilde{t}}$, and the relevant
branching fractions.  Later, by varying the parameters of the model, we
show that it is the top-squark mass itself that is the dominant variable in
probing the $\lpp$ couplings.

We begin with common scalar and fermion masses of $m_0 = 100$ GeV and
$m_{1/2} = 150$ GeV, respectively, at the grand unified theory (GUT) scale.
We choose a trilinear coupling $A_0 = -300$ GeV and the ratio of the Higgs
vacuum expectation values $\tan\beta = 4$.  The absolute value of the Higgs
mass parameter $\mu$ is fixed by electroweak symmetry breaking and is
assumed positive. Superpartner masses and decay widths are calculated with
ISAJET 7.50~\cite{Paige} and ISAWIG 1.1~\cite{HERWIG}.  After evolution
from the GUT scale to the weak scale, for this set of parameters one
obtains $m_{\widetilde{t}_1}=$ 167 GeV, $m_{\widetilde {\chi}^0_1}=$ 53
GeV, $m_{\widetilde {\chi}^{\pm}_1}=$ 95 GeV, and
$\sin\theta_{\widetilde{t}} = 0.8$.  In Fig.~\ref{fig_mcont} we show the
mass contours for the $\widetilde{t}_1$, the $\widetilde {\chi}^0_1$, and
the $\widetilde {\chi}^{\pm}_1$.  A recent experimental lower bound of
$80.5$ GeV on the mass of the lightest chargino
$\widetilde{\chi}_1^+$~\cite{WBOUND} is used to exclude regions (ex) of
small $m_{1/2}$.  Regions in which tachyonic particles would be generated,
electroweak symmetry would not be broken, or in which the lightest
supersymmetric particle (LSP) is not the lightest neutralino
$\widetilde{\chi}_1^0$ are marked as excluded theoretically (th).

The top-squark mass grows slowly with $m_0$. The lightest chargino
$\widetilde{\chi}_1^+$ and neutralino $\widetilde{\chi}_1^0$ masses are
nearly independent of the common scalar mass $m_0$.  In order to focus
principally on top-squark mass dependence, we vary $m_0$ and keep the other
supersymmetric parameters fixed.  Since the gaugino masses depend primarily
on the choice of $m_{1/2}$, variation of $m_0$ allows us to vary
$m_{\widetilde t_1}$ without an appreciable change in the masses of the
decay products and only a slow rise of $\sin\theta_{\widetilde{t}}$ to 1.
The results obtained in this section therefore depend primarily on the
top-squark mass and on which decay channels are allowed kinematically.

We reconstruct a longitudinally invariant mass $M \equiv M_T + {\not
\!E}_T$, where $M_T = \sqrt{m_X^2 + p_{T X}^2}$ is the transverse mass for
the lepton--$b$-jet system ($m_X^2=P_X^2$, $P^\mu_X = P^\mu_b +
P^\mu_l$), and ${\not \!E}_T$ is the magnitude of the missing transverse
energy for the event.  The mass $M$ has the useful feature that it produces
a peak at $m_{\widetilde{t}_1}-m_{\widetilde{\chi}_1^0}+m_b$.
Reconstruction of the peak provides a measure of the mass difference
between the top squark and the lightest neutralino.  We replace the mass
definition of Ref.~\cite{bhsletter} because it is not longitudinally
boost-invariant and hence is more sensitive to the detailed modeling of the
decays.  An added benefit is that the reach in mass with run I data is
improved.

In Fig.~\ref{fig_sba} we show a representative reconstructed mass $M$
distribution at run II of the Tevatron ($\sqrt{S}=2$ TeV) for a top squark
of mass $m_{\widetilde{t}_1}=255$ GeV.  The coupling
$\lambda^{\prime\prime}=0.1$ is chosen to be one order-of-magnitude below
the current bound.  The total background ($B$) is shown with its
components: $W j$ includes $W c$, $W j$, $W b \bar b$, and $W c \bar c$;
$t$ includes all single-top-quark production modes.  Both the ``lepton
veto'' and ``jet veto'' are applied in this figure.  The decrease in the
background for $M < 140$ GeV evident in the figure is attributed to the
cuts we impose.

It should be easy to observe the sizeable deviation from the background
distribution associated with a top squark produced with coupling strength
$\lambda^{\prime\prime}=0.1$, shown in Fig.~\ref{fig_sba}.  In
Fig.~\ref{fig_sbb} we show the distribution in $M$ for the minimum value of
$\lambda^{\prime\prime}=0.04$ required for a $5\sigma$ discovery with an
integrated luminosity of 2~fb$^{-1}$ at $\sqrt{S} = 2$ TeV.  The
significance is calculated in a bin of width $\pm15\%$ about the center of
the peak in $M$.  A bin width of $\pm10\%$ or $\pm25\%$ produces the same
significance to within a few percent.  In this plot,
$m_{\widetilde{t}_1}-m_{\widetilde{\chi}_1^0} \simeq 201$ GeV.  When this
difference is reduced to $\sim 150$ GeV, the signal and background spectra
peak in the same location, and sensitivity to the signal begins to be lost.

Performing the same mass reconstruction for a series of top-squark masses,
we examine quantitatively the functional dependence of
Eq.~(\ref{signalsigma}) on $\lambda^{\prime\prime}$.  For most of MSUGRA
parameter space, $a\Gamma_{\widetilde{t}_1}^{R_p} = 0.01$, and
$b\Gamma_{\widetilde{\chi}_1^+}^{R_p} = 0.5$.  The resulting significance
($S/\sqrt{B}$) as a function of $\lpp$ is shown in Figs. \ref{fig_siglama}
and \ref{fig_siglam} for each run of the Tevatron.  For $\lpp < 0.2$, the
significance is proportional to $(\lpp)^2/\sqrt{B}$.  This behavior has two
important consequences.  In order to improve the bounds on the $\lpp$
couplings by a further factor of 2 at run II, an additional factor of 16 in
integrated luminosity would be required.  Thus, it will be difficult to set
better limits than those presented here.  On the other hand, the full run
II integrated luminosity is not necessary to approach the limits.  Second,
the calculation is stable against uncertainties in the background
estimation, or higher order corrections to the signal.  Even a factor of 2
uncertainty in the background would not shift the limits by more than
$20\%$.  If $\lpp \approx 1$, the loss of signal is important only for the
highest masses we probe.  This region might be covered by looking for a
peak at large $E_{Tb}$ in the many-jet plus $b$-tag sample, where the
chargino has decayed to three jets.

In Fig.~\ref{fig_lammt} we show the reach in $\lambda^{\prime\prime}$ for
$165 < m_{\widetilde{t}_1} < 550$ GeV for run I and run II of the Tevatron,
and with the first 10 fb$^{-1}$ at the LHC.  With an integrated luminosity
of 2~fb$^{-1}$ at $\sqrt{S} = 2$ TeV, discovery at the level of $5\sigma$
is possible provided that $\lambda^{\prime\prime} >$ 0.02--0.1 for top
squarks with $165 < m_{\widetilde{t}_1} < 400$ GeV.  Otherwise, a $95\%$
confidence-level exclusion limit can be set for $\lambda^{\prime\prime}$ as
a function of the mass out to about 500 GeV.  For the lower integrated
luminosity and energy of the existing run~I data, values of
$\lambda^{\prime\prime} >$ 0.04--0.3 can be excluded at the $95\%$
confidence level if $m_{\widetilde{t}_1} =$ 165--350 GeV.  We note that
the run I results in Fig.~\ref{fig_lammt} represent a significant
improvement over those in our Letter \cite{bhsletter}.  The principal
reason for this change is use of the new mass variable $M$.  With the large
increase in cross section at the LHC, both discovery regions and limits may
be extended another order of magnitude in $\lpp$, and to top-squark masses
approaching 1 TeV.

There are two reasons our analysis cannot be extended much below
$m_{\widetilde{t}_1} \simeq$ 165 GeV.  One is the loss of signal owing to
cuts. While the cuts might be relaxed slightly, they are now close to the
limit.  The other is a loss of confidence in modeling of the background
near and below the peak in the $W - j$ mass distribution.  Below this peak,
mismeasurement problems and detector effects are significant sources of
uncertainty.  A detailed experimental analysis may be able to overcome this
constraint.

The virtue of probing the $\lpp_{3jk}$ couplings directly is that they may
be bounded independently if no signal is found.  The limits on individual
couplings are determined by the relative weights of the corresponding parton
luminosities.  In Figs. \ref{fig_lammtea} and \ref{fig_lammteb}, we show
the $95\%$ confidence-level limits that may be placed on $\lpp_{312}$,
$\lpp_{313}$, and $\lpp_{323}$ independently, as a function of top-squark
mass at run I and II of the Tevatron.  For $m_{\widetilde{t}_1}= 255$ GeV,
absence of a signal in run I data means that $\lpp_{312} < 0.08$, and
absence in run II data means that $\lpp_{312} < 0.03$.  The bound on
$\lpp_{312}$ is strongest because its contribution is proportional to the
$d \otimes s$ parton luminosity.  Even at run I, a significant improvement
can be made in the limits on $\lpp_{312}$ and $\lpp_{313}$.

Another aspect of Eq.~(\ref{signalsigma}) is apparent in the crossing of
the limits for a common $\lpp$ and $\lpp_{312}$.  The three possible
couplings contribute almost equally to the \RPV decays of the top squark
and chargino.  In contrast, the production rate arises mostly from the
$\lpp_{312}$ term.  Hence, there is a net gain in measurable cross section
in the $R_p$-conserving decay channel if only this coupling is large.  At
run II, a limit of $\lpp_{312} < 0.2$ can be placed for top-squark masses
up to 550 GeV.  Even $\lpp_{323}$ may be probed directly if
$m_{\widetilde{t}_1} < 300$ GeV.

\subsection{MSUGRA}
\label{sugraresults}

A specific set of parameters is used in the prior section to generate the
masses of the top squark and its decay products.  In this section we undertake
a broader examination of the minimal supergravity model (MSUGRA) parameter
space, and we present limits as a function of $m_0$ and $m_{1/2}$ for various
choices of $A_0$, $\tan{\beta}$, and the sign of $\mu$.  This study allows us
to conclude that for most of MSUGRA parameter space the top-squark discovery
potential and bounds depend most strongly on the the top-squark mass alone.

\begin{figure}[tb]
\begin{center}
\epsfxsize= 3.26in 
\leavevmode
\epsfbox{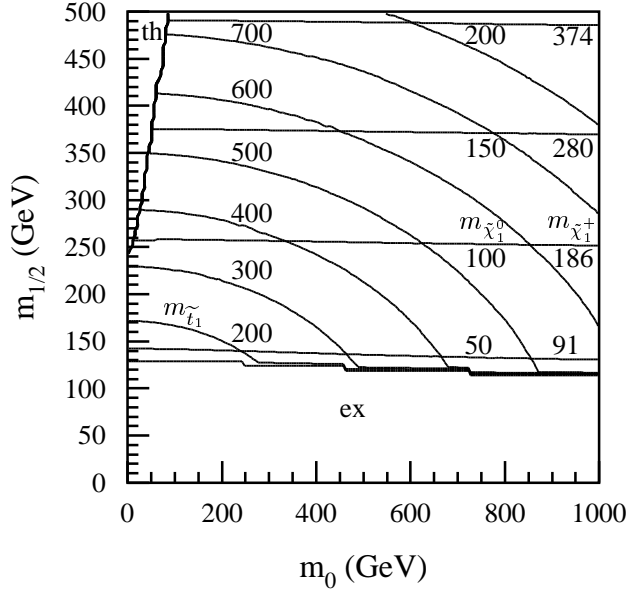}
\end{center}
\caption{Contours of constant top-squark mass $m_{\widetilde{t}_1}$,
chargino mass $m_{\widetilde{\chi}_1^+}$ and neutralino mass
$m_{\widetilde{\chi}_1^0}$ (GeV) through the $m_0 - m_{1/2}$ plane
in MSUGRA with $\tan{\beta} = 4$, $A_0 = -300$ GeV, and $\mu > 0$.
Theoretically excluded regions are indicated with (th) and
experimentally excluded regions with (ex).}
\label{fig_mcont}
\end{figure}

\begin{figure}[tb]
\begin{center}
\epsfxsize= 3.0in 
\leavevmode
\epsfbox{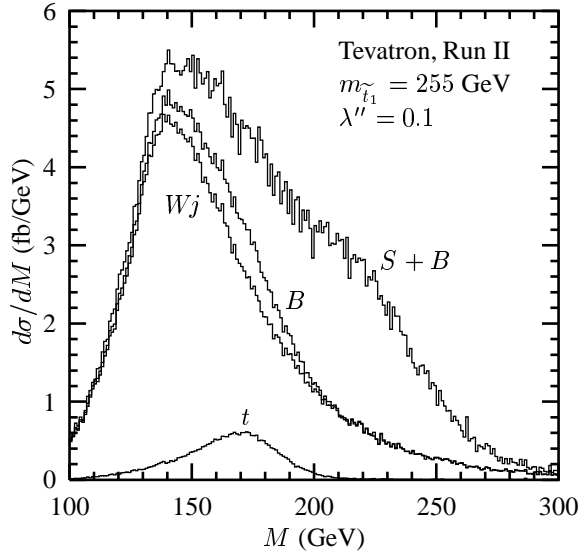}
\end{center}
\caption{The reconstructed mass $M$ distribution for single-top-squark
production ($S$) and backgrounds ($B$) at the Tevatron ($\sqrt{S}=2$
TeV) for a top-squark mass $m_{\widetilde{t}_1}=255$ GeV and coupling
$\lambda^{\prime\prime}=0.1$ with all cuts applied. The $W j$
component of the background includes $W c$, $W j$, $W b \bar b$, and
$W c \bar c$.  The $t$ component of the background includes all
single-top-quark production modes ($t\bar b j$ is dominant; $t\bar b$ and
$W t$ are negligible).}
\label{fig_sba}
\end{figure}

\begin{figure}[tb]
\begin{center}
\epsfxsize= 3.0in 
\leavevmode
\epsfbox{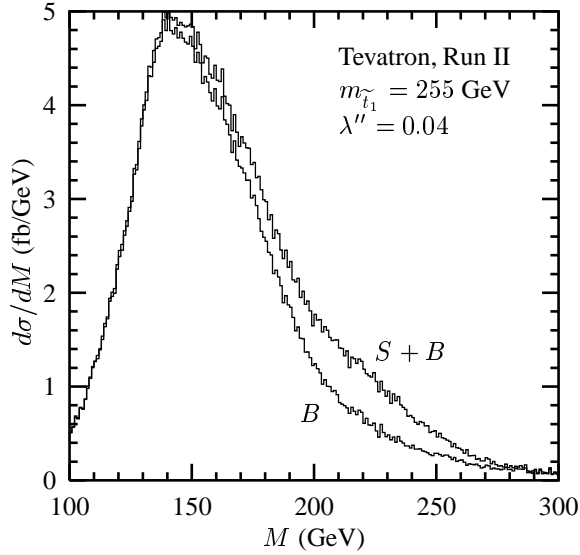}
\end{center}
\caption{The reconstructed mass $M$ distribution for single-top-squark
production ($S$) and backgrounds ($B$) at the Tevatron ($\sqrt{S}=2$
TeV) for a top-squark mass $m_{\widetilde{t}_1}=255$ GeV with all cuts
applied.  The coupling $\lambda^{\prime\prime}=0.04$ produces the
minimum signal for a $5\sigma$ significance at this mass.}
\label{fig_sbb}
\end{figure}

\begin{figure}[tb]
\begin{center}
\epsfxsize= 2.85in 
\leavevmode
\epsfbox{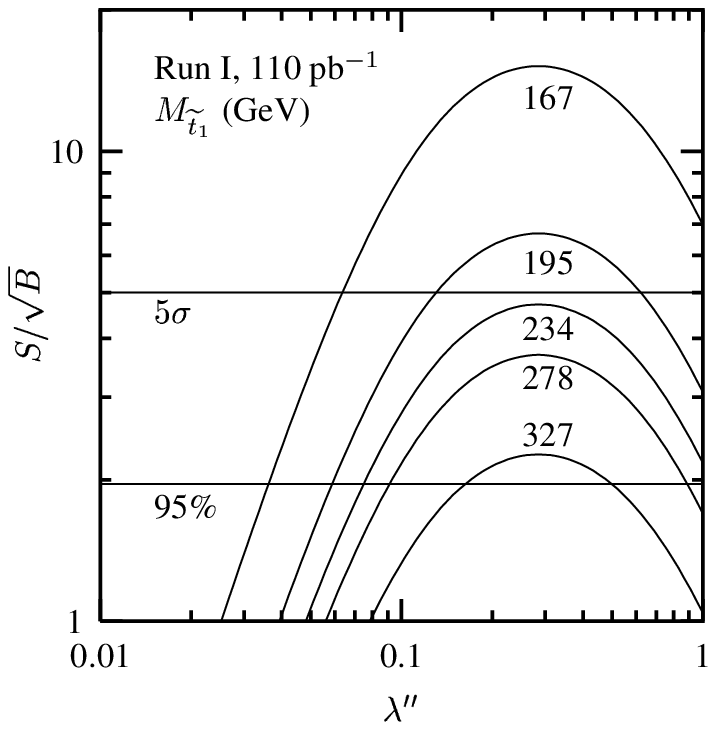}
\end{center}
\caption{Statistical significance of the single-top-squark signal
($S/\sqrt{B}$) in run~I of the Tevatron ($\sqrt{S}=1.8$ TeV, 110 pb$^{-1}$)
versus $\lambda^{\prime\prime}$ for a variety of top-squark masses (GeV).}
\label{fig_siglama}
\end{figure}

\begin{figure}[tb]
\begin{center}
\epsfxsize= 2.85in 
\leavevmode
\epsfbox{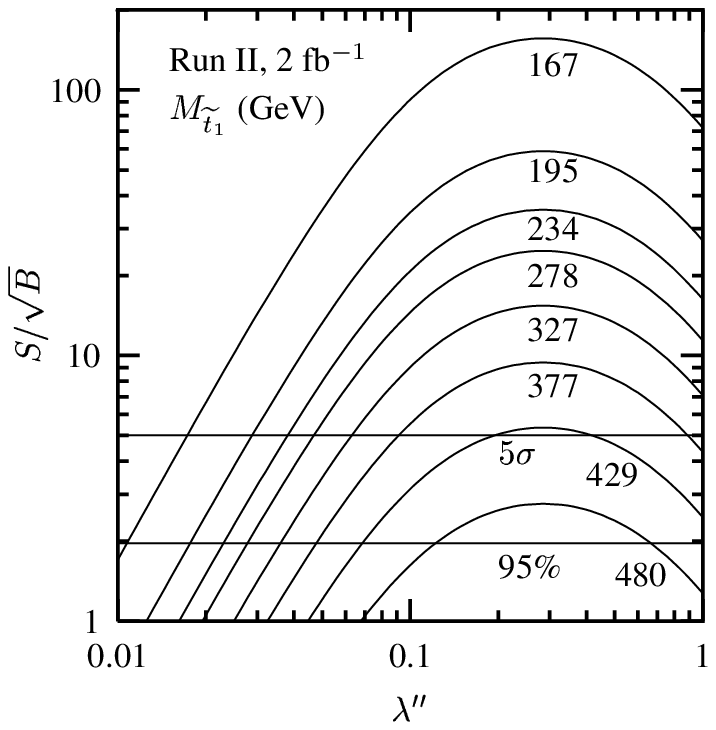}
\end{center}
\caption{Statistical significance of the single-top-squark signal
($S/\sqrt{B}$) in run~II of the Tevatron ($\sqrt{S}=2$ TeV, 2 fb$^{-1}$)
versus $\lambda^{\prime\prime}$ for a variety of top-squark masses (GeV).}
\label{fig_siglam}
\end{figure}

\begin{figure}[tb]
\begin{center}
\epsfxsize= 2.86in 
\leavevmode
\epsfbox{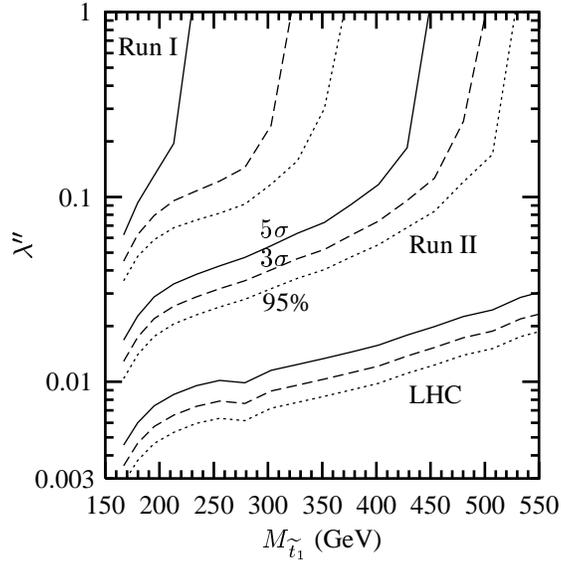}
\end{center}
\caption{Lower limits on discovery ($S/\sqrt{B} = 5$, solid), evidence
($S/\sqrt{B} = 3$, dashed), and 95$\%$ confidence-level exclusion
($S/\sqrt{B} = 1.96$, dotted) for $\lambda^{\prime\prime}$ versus
top-squark mass in run~I of the Tevatron ($\sqrt{S}=1.8$ TeV,
110~pb$^{-1}$), run~II of the Tevatron ($\sqrt{S}=2$ TeV,
2~fb$^{-1}$), and one year at the LHC ($\sqrt{S}=14$ TeV,
10~fb$^{-1}$).}
\label{fig_lammt}
\end{figure}

\begin{figure}[tb]
\begin{center}
\epsfxsize= 2.86in 
\leavevmode
\epsfbox{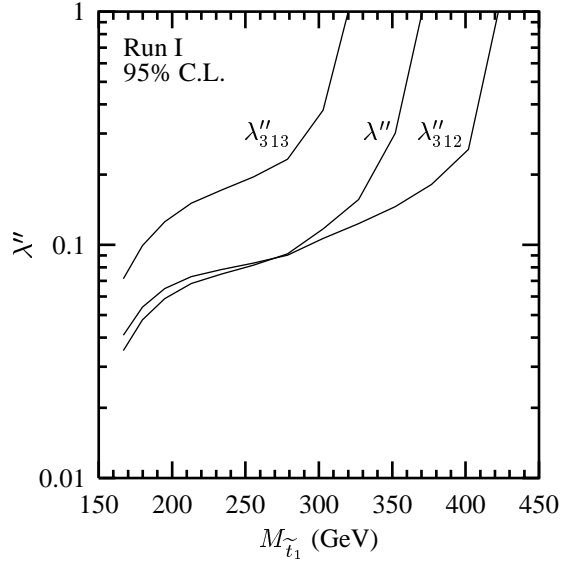}
\end{center}
\caption{95$\%$ confidence-level exclusion limits ($S/\sqrt{B} =
1.96$) for $\lambda^{\prime\prime}_{3jk}$ versus top-squark mass in
run~I of the Tevatron ($\sqrt{S}=1.8$ TeV, 110~pb$^{-1}$).  The curve
marked by $\lambda^{\prime\prime}$ is obtained if all three
$\lambda^{\prime\prime}_{3jk}$ are set equal.  There is insufficient rate
to obtain a 95$\%$ limit on $\lambda^{\prime\prime}_{323}$ at run I.}
\label{fig_lammtea}
\end{figure}

\begin{figure}[tb]
\begin{center}
\epsfxsize= 2.86in 
\leavevmode
\epsfbox{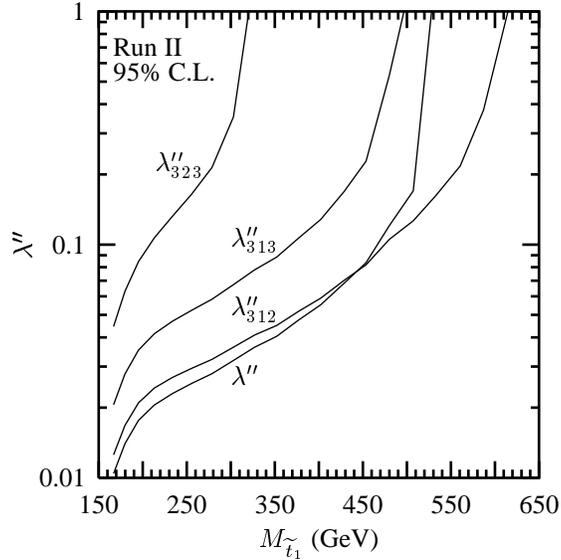}
\end{center}
\caption{95$\%$ confidence-level exclusion limits ($S/\sqrt{B} =
1.96$) for $\lambda^{\prime\prime}_{3jk}$ versus top-squark mass in
run~II of the Tevatron ($\sqrt{S}=2$ TeV, 2~fb$^{-1}$).}
\label{fig_lammteb}
\end{figure}

\begin{figure}[tb]
\begin{center}
\epsfxsize= 6.5in 
\leavevmode
\epsfbox{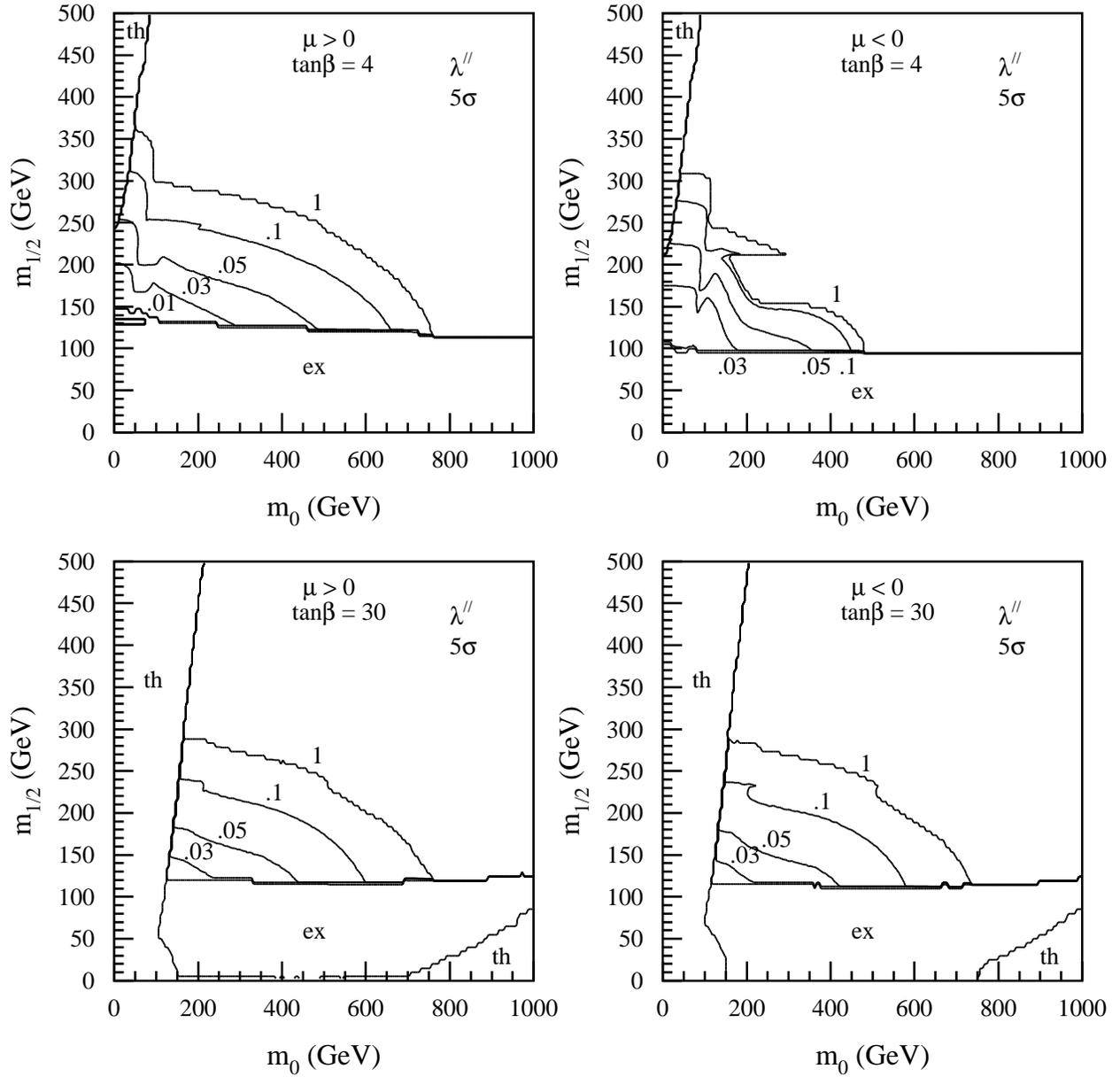}
\end{center}
\caption{Contours of the minimum $\lambda^{\prime\prime}$ needed for a
($5\sigma$) discovery with 2 fb$^{-1}$ of data at run II of the
Tevatron.  Four $m_0 - m_{1/2}$ MSUGRA planes are shown with $A_0 =
-300$ GeV, $\mu > 0$ or $\mu < 0$, and $\tan{\beta} = 4$ or $\tan{\beta}
= 30$.  Theoretically excluded regions are indicated with (th), and
experimentally excluded regions are indicated with (ex).  For a given
$\lambda^{\prime\prime}$, discovery should occur for values of
$m_0 - m_{1/2}$ below and left of the designated contour.}
\label{fig_sugd}
\end{figure}

In Fig.~\ref{fig_sugd} we show contours in the $m_0 - m_{1/2}$ plane of the
minimum $\lpp$ necessary for a $5\sigma$ discovery of the top squark given
2~fb$^{-1}$ of integrated luminosity at run II of the Tevatron.  Plots are
shown for $A_0 = -300$ GeV, $\tan{\beta} = 4$ and $\tan{\beta} = 30$, and
both signs of $\mu$.  The regions of experimental and theoretical exclusion
are explained above, in conjunction with Fig.~\ref{fig_mcont}.  At low
$m_0$ the $\tau\nu\widetilde{\chi}_1^0$ decay of $\widetilde{\chi}_1^+$
saturates its \RPC branching fraction.  Naively, this effect would seem to
make the electron and muon signal disappear.  However, a small fraction of
the taus satisfies the electron or muon tagging conditions.  The net effect
is an increase in the mass reach in the region of small $m_0$, most evident
in the contour plots for $\tan{\beta} = 4$.

The discovery contours of Fig.~\ref{fig_sugd} follow closely the top-squark
mass contours.  This point becomes evident if one compares
Fig.~\ref{fig_mcont} and the top-left-hand pane of Fig.~\ref{fig_sugd} for
which the same MSUGRA parameters are used.  For example, the contour for
$\lpp = 0.1$ in the top-left-pane of Fig.~\ref{fig_sugd} can be seen to
correspond to $m_{\widetilde t_1} = 350$--$375$ GeV in
Fig.~\ref{fig_mcont}.  The same statement is true also for large
$\tan{\beta}$ and both signs of $\mu$: $\lpp = 0.1$ corresponds to
$m_{\widetilde t_1} = 350$--$375$ GeV.

The only region of significant dependence on the MSUGRA parameters, beyond
those that set the top-squark mass, is at low $\tan{\beta}$ and $\mu <
0$. There is a large suppression of the leptonic branching fractions of the
chargino when $|\mu | \sim m_{1/2} \sim m_0$.  This suppression is due
mostly to a destructive interference between the $W$ boson and sneutrino
($\widetilde \nu$) decay modes in this region~\cite{BFMM}.  In terms of
physical parameters, $m_{\widetilde \nu}$ is close to $|\mu |$, and the sign of
the $W - \widetilde{\nu}$ interference term is the same as the sign of
$\mu$.  Hence only the region of negative $\mu$ is affected.  However, the
cancellation decouples, as expected, as $m_{\widetilde \nu}$ becomes heavier
than $m_W$.  Dependence of the bounds on the trilinear coupling $A_0$ is
negligible except in the region of small $\tan \beta$ and $\mu < 0$ where
it serves as a moderator of the cancellation.  As $A_0$ becomes more
positive, the cancellation increases.

The results in Fig.~\ref{fig_sugd} demonstrate that the reach in top-squark
mass is significantly larger in single-top-squark production than it is for
pair production, where the latter is restricted to the lower left-hand
corner of the plot ($m_0 < 300$ GeV, $m_{1/2}<170$ GeV)~\cite{msugrarep}.

\begin{figure}[tb]
\begin{center}
\epsfxsize= 6.5in 
\leavevmode
\epsfbox{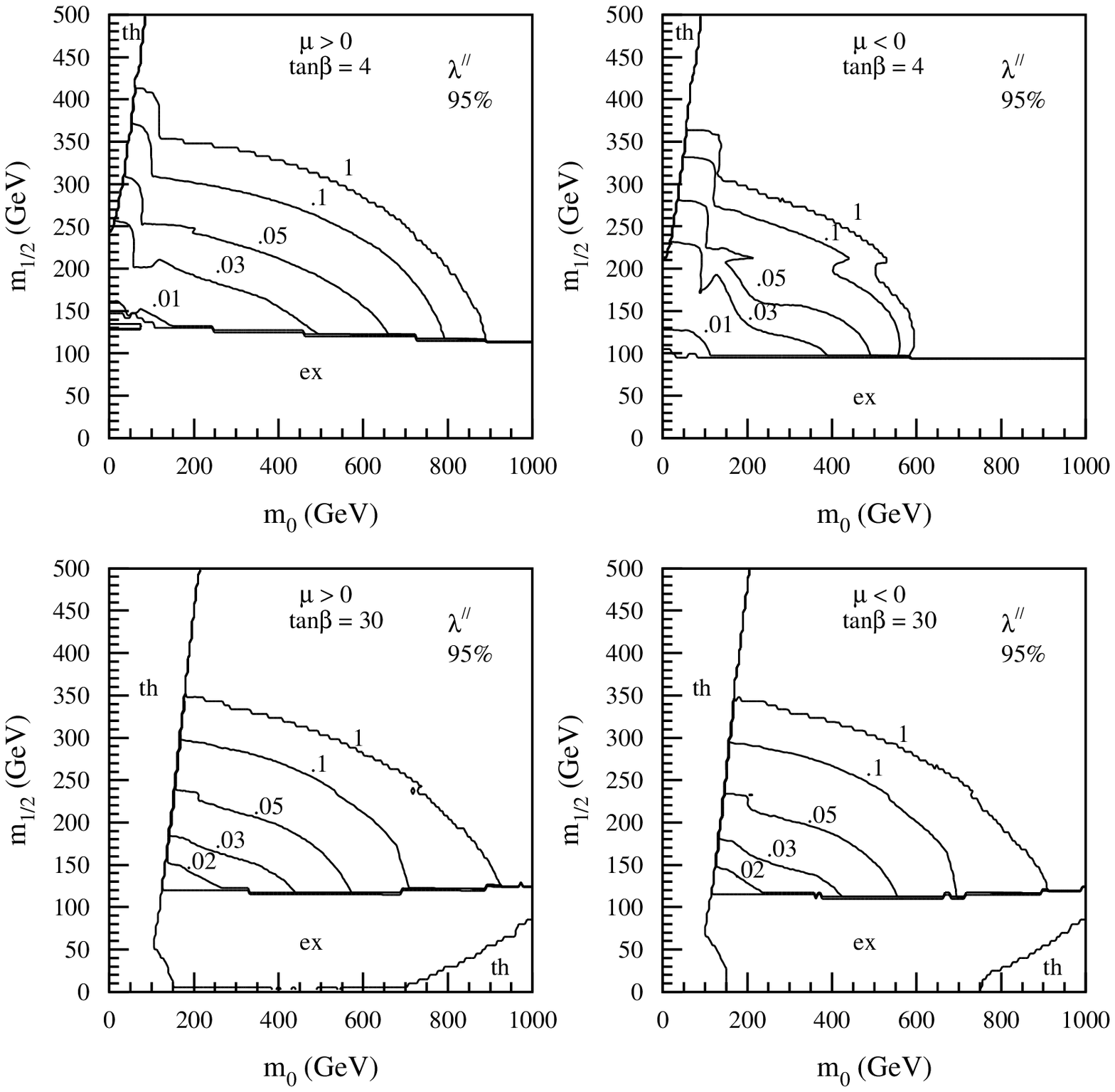}
\end{center}
\caption{$95\%$ confidence-level ($1.96\sigma$) exclusion contours that can
be set for $\lambda^{\prime\prime}$ with 2 fb$^{-1}$ of data at run II of
the Tevatron.  Four $m_0 - m_{1/2}$ MSUGRA planes are shown with $A_0 =
-300$ GeV, $\mu > 0$ or $\mu < 0$, and $\tan{\beta} = 4$ or $\tan{\beta} = 30$.
Theoretically excluded regions are indicated with (th), and experimentally
excluded regions are indicated with (ex).  Absence of a signal excludes
values of $\lpp$ in the region below and left of the designated contour.}
\label{fig_suge}
\end{figure}

Contours in the $m_0 - m_{1/2}$ plane of the $95\%$ confidence-level limits
($1.96\sigma$) are shown in Fig.~\ref{fig_suge} for the same parameters as
in Fig.~\ref{fig_sugd}.  Except for the same region of suppression at low
$\tan\beta$ and $\mu < 0$, limits can be improved with data from run II at
the Tevatron by at least an order-of-magnitude for a broad range of $m_0 <
800$ GeV and $m_{1/2} < 300$ GeV.  The exclusion contours of
Fig.~\ref{fig_suge} also follow closely the top-squark mass contours.  The
contour for $\lpp = 0.1$ in the top-left-pane of Fig.~\ref{fig_suge} can be
seen to correspond to $m_{\widetilde t_1} = 450$ GeV in
Fig.~\ref{fig_mcont}.  This remains true for large $\tan{\beta}$ and both
signs of $\mu$: $\lpp = 0.1$ corresponds to $m_{\widetilde t_1} = 450$ GeV.

\begin{figure}[tb]
\begin{center}
\epsfxsize= 6.5in 
\leavevmode
\epsfbox{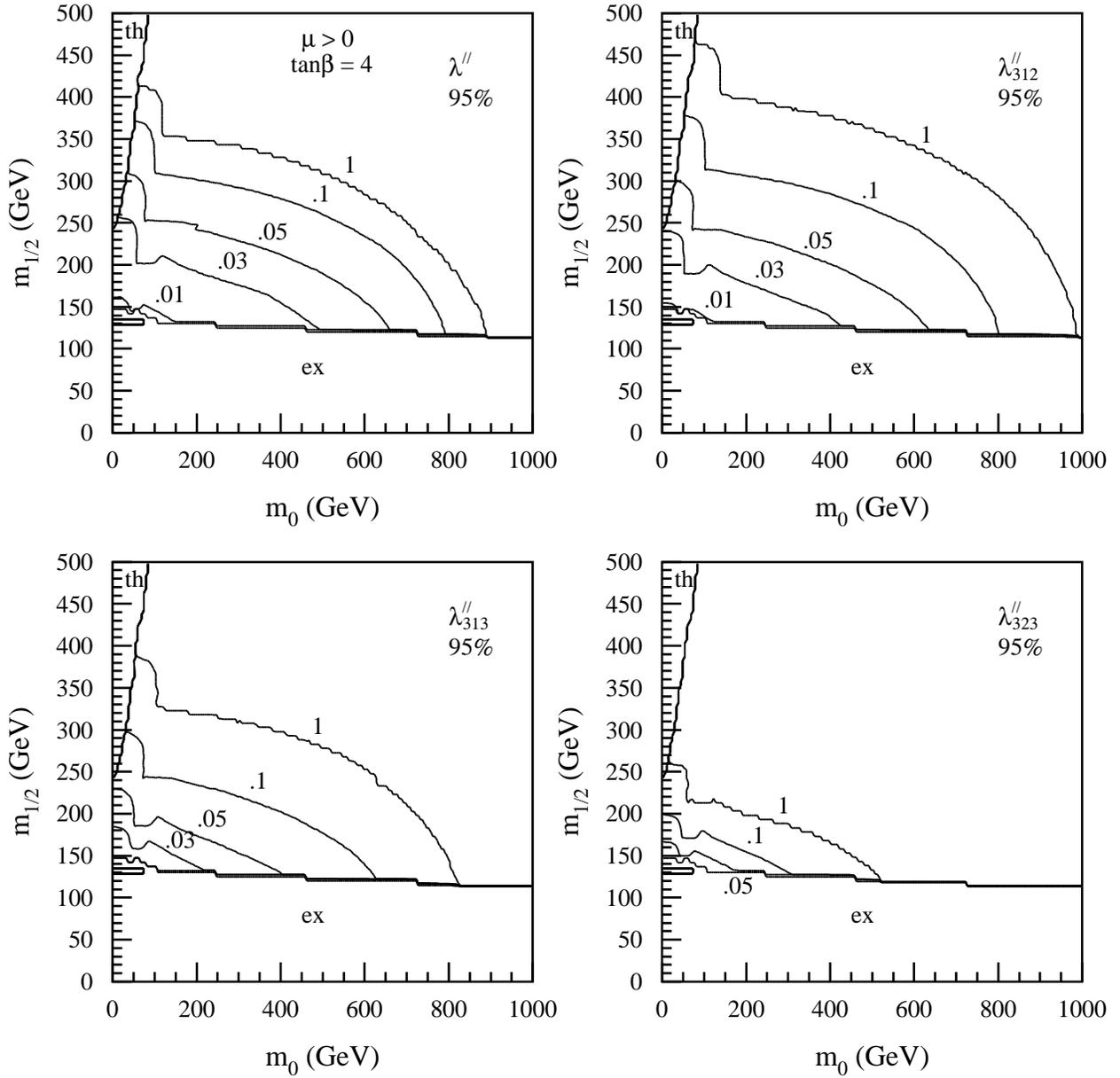}
\end{center}
\caption{$95\%$ confidence-level ($1.96\sigma$) exclusion contours that can
be set for $\lambda^{\prime\prime}_{312}$, $\lambda^{\prime\prime}_{313}$,
or $\lambda^{\prime\prime}_{323}$, with 2 fb$^{-1}$ of data at run II of
the Tevatron.  Four $m_0 - m_{1/2}$ MSUGRA planes are shown with $A_0 =
-300$ GeV, $\mu > 0$, and $\tan{\beta} = 4$.  Theoretically excluded regions
are indicated with (th), and experimentally excluded regions are indicated
with (ex).  Absence of a signal excludes values of the $\lpp_{3jk}$
couplings in the region below and left of the designated contour.}
\label{fig_sugjk}
\end{figure}

The strength of a direct probe of the \RPV couplings is that limits may be
extracted for each $\lpp_{3jk}$ coupling separately.  In
Fig.~\ref{fig_sugjk}, we present contours in the $m_0 - m_{1/2}$ plane of
the $95\%$ confidence-level limits that may be placed on $\lpp_{312}$,
$\lpp_{313}$, or $\lpp_{323}$ on the assumption that only one coupling is
non-negligible.  The $\tan\beta = 4$, $\mu > 0$ limits for a common $\lpp$
from Fig.~\ref{fig_suge} are also shown for comparison.  Contours for each
$\lpp_{3jk}$ follow contours of constant top-squark mass just as for a
common $\lpp$.  Hence, the top-squark masses corresponding to $\lpp_{312}$,
$\lpp_{313}$, or $\lpp_{323} = 0.1$ are $m_{\widetilde t_1} \approx 450$
GeV, $350$ GeV, or $215$ GeV, respectively.  In terms of a minimal
supergravity model, the limits achievable at the Tevatron on individual
$\lpp_{3jk}$ couplings can be estimated for any value of $m_0$ and
$m_{1/2}$ by comparing Fig.~\ref{fig_mcont} and Fig.~\ref{fig_lammtea} or
Fig.~\ref{fig_lammteb}.


\section{Summary}
\label{summary}

In this paper, we extend our study of $s$-channel production of a
{\it{single}} top squark in hadron collisions through an $R_p$-violating
mechanism~\cite{bhsletter}.  Even if the $R_p$-violating couplings
$\lambda^{\prime\prime}$ are as small as $0.01$, two orders of magnitude
below the current bounds, the single-top-squark rate exceeds the pair rate
for $m_{\widetilde{t}_1} > 100$ GeV.  This enhancement will allow searches
at the Fermilab Tevatron to reach much higher top-squark masses than those
based on traditional squark pair-production.

Because of the overwhelming jet rate from standard strong interaction
processes, \RPV production of top squarks followed by \RPV decay into a
pair of jets is not a viable means to exclude $R$-parity-violating
production of top squarks or to obtain useful bounds on $m_{\widetilde
t_1}$ or $\lambda^{\prime\prime}$.

We study observability of single top squarks by focusing on the clean
$R_p$-conserving decay, $\widetilde{t}_1 \rightarrow b
\widetilde{\chi}^+_1$, with $\widetilde{\chi}^+_1 \rightarrow l^+ + \nu_l
+\widetilde{\chi}^0_1$.  The combined branching fraction for this cascade
decay is of order 20\% or greater for $\lambda^{\prime\prime} <$ 0.1, a
region of significant interest.  We address the case where the neutralino
either decays outside of the detector volume or leaves remnants that are
too soft to identify; its signature is missing energy in the detector.  We
simulate both the signal and standard model background processes, including
parton showering and hadronization as well as a full detector simulation.
We investigate sensitivity to parameters in our analysis by exploring a
wide range of the parameter space of the minimal supergravity model of
supersymmetry breaking.

With an integrated luminosity of 2~fb$^{-1}$ at $\sqrt{S} = 2$ TeV, we show
that discovery of a top squark with $m_{\widetilde{t}_1} < 400$ GeV at the
level of $5\sigma$ is possible provided that the \RPV coupling
$\lambda^{\prime\prime} >$ 0.02--0.1.  Otherwise, a $95\%$
confidence-level exclusion limit can be set for $\lambda^{\prime\prime}$ as
a function of the mass out to about 500 GeV.  For the lower integrated
luminosity and energy of the existing run~I data, values of
$\lambda^{\prime\prime} >$ 0.05--0.2 can be excluded at the $95\%$
confidence level if $m_{\widetilde{t}_1}=$ 165--350 GeV.  These limits
would constitute a significant improvement over the current $95\%$
confidence-level upper bounds of $\lambda^{\prime\prime} \lesssim 1$ for
squarks of the third generation. With the large increase in cross section
at the CERN Large Hadron Collider, the discovery region and limits may be
improved by another order of magnitude in $\lpp$ and extended to top-squark
masses approaching 1 TeV.  The discovery reach, or exclusion limits, at the
Tevatron and LHC depend almost exclusively on the value of the top-squark
mass, largely independent of other MSUGRA parameters.  Since the $R$-parity
violation occurs only at the production stage, the $\lpp_{3jk}$ couplings
may be bounded independently if no signal is found.  Even at run I, a
significant improvement can be made in the limits on $\lpp_{312}$ and
$\lpp_{313}$.

We propose a simple method to directly probe $R$-parity-violating couplings
at hadron colliders.  This analysis searches for a single top squark that
is produced in the $s$-channel via a baryon-number-violating interaction,
and decays via standard $R$-parity-conserving interactions.  This strategy
should be extended to other squarks and $R$-parity-violating couplings.


\section*{Acknowledgments}
We thank Tom LeCompte and Steve Kuhlmann for valuable advice.  We
acknowledge communications with Peter Richardson regarding the use of
ISAWIG and with Tilman Plehn regarding the use of his NLO $K$-factors.
This work was supported by the U.S. Department of Energy, High Energy
Physics Division, under Contract No. W-31-109-Eng-38.



\begin{table}
\caption{Cuts used to simulate the acceptance of the detector for the
Tevatron run~II and run~I (in parentheses if different). At the LHC,
$|\eta|<2.5$ for all objects. \label{tab_acc}}
\begin{center}
\begin{tabular}{ll}
$|\eta_b|<2$ (1) & $E_{Tb}>40$ GeV \\
$|\eta_l|<1.5$ (1.1) & $E_{Tl}>15$ GeV (20 GeV)\\
$|\eta_j|<2.5$ & $E_{Tj}>30$ GeV\\
${\not \!E}_{T}>20$ GeV &
\end{tabular} \end{center}
\end{table}

\end{document}